# Application Of Module to Coding Theory: A Systematic Literature Review

**Muhammad Faldiyan, Sisilia Sylviani**

Department of Mathematics, Faculty of Mathematics and Natural Sciences, Padjadjaran University, Indonesia

\*        Correspondence: sisilia.sylviani@unpad.ac.id

## ABSTRACT

A systematic literature review is a research process that identifies, evaluates, and interprets all relevant study findings connected to specific research questions, topics, or phenomena of interest. In this work, a thorough review of the literature on the issue of the link between module structure and coding theory was done. A literature search yielded 470 articles from the Google Scholar, Dimensions, and Science Direct databases. After further article selection process, 14 articles were chosen to be studied in further depth. The items retrieved were from the previous ten years, from 2012 to 2022. The PRISMA analytical approach and bibliometric analysis were employed in this investigation. A more detailed description of the PRISMA technique and the significance of the bibliometric analysis is provided. The findings of this study are presented in the form of brief summaries of the 14 articles and research recommendations. At the end of the study, recommendations for future development of the code structure utilized in the articles that are further investigated are made.

**Keywords**: module structure, coding theory, linear code,

## INTRODUCTION

A module over a ring is a commutative algebra over commutative ring. The study of commutative rings and related structures, particularly ideal and modules, is known as commutative algebra [1]. The module structure in the ring domain is the same as the multiplication and addition operations in the ring [2]. A module structure over ring $A$ on $A^n$ is denoted by addition operations i.e $x + y = (x_1 + y_1, \dots, x_n + y_n)$ and multiplication operations i.e $ax = (ax_1, \dots, ax_n)$, for $x = (x_1, \dots, x_n), y = (y_1, \dots, y_n) \in A^n$ and for $a \in A$ [3]. Module is an algebraic structure which means it also has submodules, bimodules, homomorphisms, and isomorphisms. Let $N$ is a non-empty set which is a subset of module $R$ is said to be a submodule of module R if $N$ is a subgroup of $R$ for the addition operation and $N$ is closed for the multiplication operation [4]. Fix an $R$-module $Q \subseteq R^n$ and define $M(Q)$ to designate the collection of $R$-submodules of $Q$. Let $M, N \in M(Q)$, the smallest submodule of $Q$ that contains both $M, N$. When $M \cap N = 0$, write $M \oplus N$. Let $Q$ and its submodules are also finite since $R$ is finite [5]. Let $M$ and $N$ be $R - S$ —bimodules and $S - T$ —bimodules, respectively, and $R, S, T$ be unital (potentially non-commutative) rings. Following that, an $R - T$ —bimodule is formed by its tensor products $M \otimes_S N$ in the normal manner. Take into consideration the canonical $S$-bimodule $R$ with actions $\rho(s)r = sr, rs = r\rho(s), \forall r \in R$ and $\forall s \in S$, for a homomorphism of unital rings $\rho: S \to R$. $R$ obviously becomes a $R - S$ —bimodule as well as a $S - R$ —bimodule [6]. There are many more concepts related to modules, because module are broad algebraic structures and can be applied anywhere, including in coding theory.

Modules can be applied in coding theory, especially in linear code. Ring-linear coding theory is an area of algebraic coding theory where the underlying alphabet just carries the structure of a finite ring or, more broadly, a module rather than the structure of a finite field [7]. An $A$-code, also known as a linear code $C$ of length $l$ over the ring $A$, is a submodule of $B = A^l$ where $C$ is stable upon addition with regard to coordinates and upon scalar multiplication by any element of $A$ [8]. A left submodule $C \subset R^n$ can be defined as a left linear code over a finite ring $R$ [9]. Additive codes are a generalizing of the idea of linear codes over a finite field. An additive code of length $n$, given a finite field $F$, is a subset of $F^n$, which is a $K$-linear subspace for some subfield $K \subseteq F$. Since it is used to create quantum error-correcting codes, additive codes have lately gained more significance, but it still interesting on their own [10].

To create an efficient and ideal code word for transmitting a message, many other academics have studied the usage of module structures in coding theory. The module structures may be used as a coding framework for transmitting messages in a wide variety of ways. Ling & Sole Research ([11]–[13]) explains about the structure of a quasi-cyclic codes which is a module structure accompanied by its use in life. Boucher et al. Research ([14], [15]) introduces skew-cyclic codes and skew-constacyclic codes in general, which are a module structure over fields. Cuadra, Garcia-Rubira, & Lopez-Ramos Research ([16], [17]) describes a code which is a module structure over Hopf Algebras, where the Hopf Algebras is made up of a linear map S:H→H termed the antipode and a bialgebra H. A bialgebra is a vector space H with compatible unital algebra structure (m,u) and co-unital co algebra structure (Δ,ε) [18]. Climent, Gluesing-Luerssen et al., and Lally Research ([19]–[21]) describes about convolutional codes which is a module structures and its applications to design any systems, create a matrix from a single code, and determine the smallest distance between codes. Sun et al. Research [22] introduces about Lattice Network Codes which is a module structure too, due to Eisenstein integers. Lam & Leroy Research ([23], [24]) explains about Wedderburn polynomial over division ring, where Wedderburn polynomial is a minimum polynomial of an algebraic subset of a division ring [25]. Reed-Solomon codes introduced by Guruswami [26], Koetter ([27]–[29]), Lee ([30], [31]), Alekhnovich [32], Zeh [33], Xing ([34], [35]), and Chen ([36], [37]) to solve coding problems such as decoding decision making, interpolation algorithm with Grobner Basis, solve polynomial of linear diophantine equations, low-complexity decoding problem, and algebraic chase decoding problem. There are still many other research involving the module structure of code, in-depth analysis of its structure, and usage in coding theory.

The purpose of this SLR research is to find strategy that will help overcome the problems encountered in compatible papers to be understood further and to identify different perspective related to the problem of coding and the structure of module. The results of this study are in the form of an analysis of the entire contents of the paper and suggestions for researchers to improve and perfect the contents of the paper. The analysis carried out involved bibliometric analysis in the search for paper literature that one wanted to understand more deeply, then followed by analysis of the paper involving the author, year of publication, number of citations, module structure used, research objectives, and research results. At the end of analysis, research suggestions will be given to researchers whose paper will be further analyzed to improved their research results. The papers that are further analyzed in the systematic literature review are Ozbudak & Ozkaya Research [38], Shi et al. Research [39], Gorla & Ravagnani Research [40], Berger & Amrani Research [41], Boulagouaz & Leroy Research [42], Torrecillas et al. Research ([43], [44]), Xing et al. Research [34], Dyshko & Wood Research [45], Guneri & Ozkaya Research [46], Morales Research [47], Borello & Willems Research [12], Westerback Research [48], Garcia-Rubira & Lopez-Ramos Research [49], and how to select the paper will be explained further in the methods section

**METHODS**

The research method contains explanations in the form of paragraphs about the research design or descriptions of the experimental settings, data sources, data collection techniques, and data analysis conducted by the researcher. This guide will explain about writing headings. If your headings exceed one, use the second level of headings as below. In this study, researchers identified and selected the literature that focuses on the use of the module structure in coding theory. The data used in this study are articles obtained from three database sources, namely Dimensions, Science Direct, and Google Scholar. The articles considered in the literature are journal articles, preprint, and pro-ceedings. The articles used for the literature review are articles published in the last ten years, namely from 2012 to 2022. Article searches were carried out using Publish or Perish software to get article data from google scholar and using database website to get article data from Dimensions and Science Direct. Article searches are limited to a maximum limit of 1000 articles on the Publish or Perish software. The keywords used by researchers to selected articles are "Module Structure", "Coding Theory" and "Alge-bra".

The Selection process for paper selection is divided into two stages, namely semi-automatic selection and manual selection. Semi-automatic selection was carried out with the help of the Jabref application to remove existing duplication from the three journal sources. Manual selection is made to read duplicates that are not detected by the Jabref application. The application is not detected due to the different file types of each journal source. Additionally, the selection of article abstracts, accessibility, and material that satisfies the author's search criteria is done manually. The manual selection process aids researchers in accurately representing research results and ensures that no studies are missed as a result of machine errors, while the semi-automatic selection process speeds up the search for duplications or increases researchers' efficiency when con-ducting literature searches.

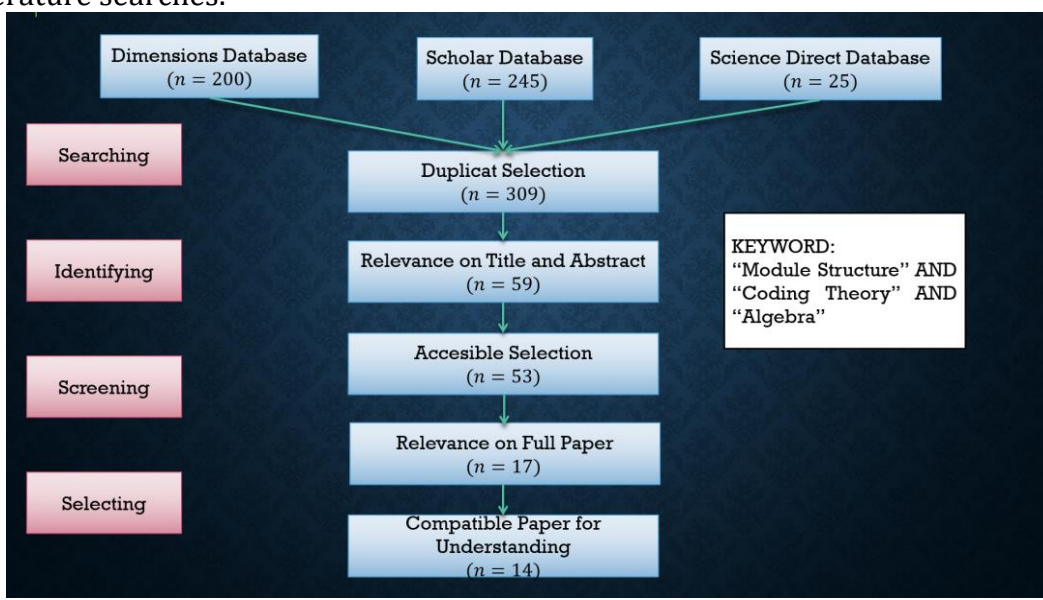

Figure 1. Flowchart of literature selection process

The following is a description of the steps involved in Figure 1. First, the literature is searched using publish or perish software by entering keywords, year of publication, and maximum article limit. Second, the literature obtained was 245 articles from the Publish or Perish software which were selected by eliminating literature in the form of books and other topics that were not relevant to this research. The selected literature is journal articles, preprint, and proceedings whose topics are relevant to module structure in coding theory. Third, open the dimensions website and search for literature by entering keywords, year of publication, and removing literature in the form of books in literature search. Findings from a literature search on the Dimensions website revealed up to 200 article that were pertinent to the issue. Fourth, open the science direct website and search for literature by entering keywords, year of publication, and removing literature in the form of books in literature search. The results of a literature search on the Science Direct Website found as many as 25 article that were relevant to the topic being searched. Fifth, obtained total 470 articles from dimensions, google scholar, and science direct which were then identified in all of these articles to eliminate duplicate articles from the three databases. Sixth, after deleting duplicate articles, 309 articles were ac-quired. These articles were chosen using the Jabref application and manual selection was done for data with various file extensions. Seventh, obtained 59 articles whose titles are relevant to the keyword entered. Reading the abstracts and titles from the previous results, then choosing the ones that come the closest to the topic's appropriateness, is how titles and abstracts that are pertinent to the topic are chosen. Eighth, with accessi-ble selection, obtained 53 articles that can be accessed for free by researchers. Searching for open sources paper, fully accessible articles is how the accessibility selection process is carried out. Nineth, read the 17 articles and select the relevant articles in full paper. Full paper selection is done by reading research results and research conclusions from previously obtained articles. Last, select compatible articles for further understanding by researchers and obtained 14 compatible articles which were selected in a systematic literature review for massive understanding. This selection is carried out by reading the paper from the introduction to conclusion and choosing those with a good research index. The research topics mapping is shown in Figure 2.

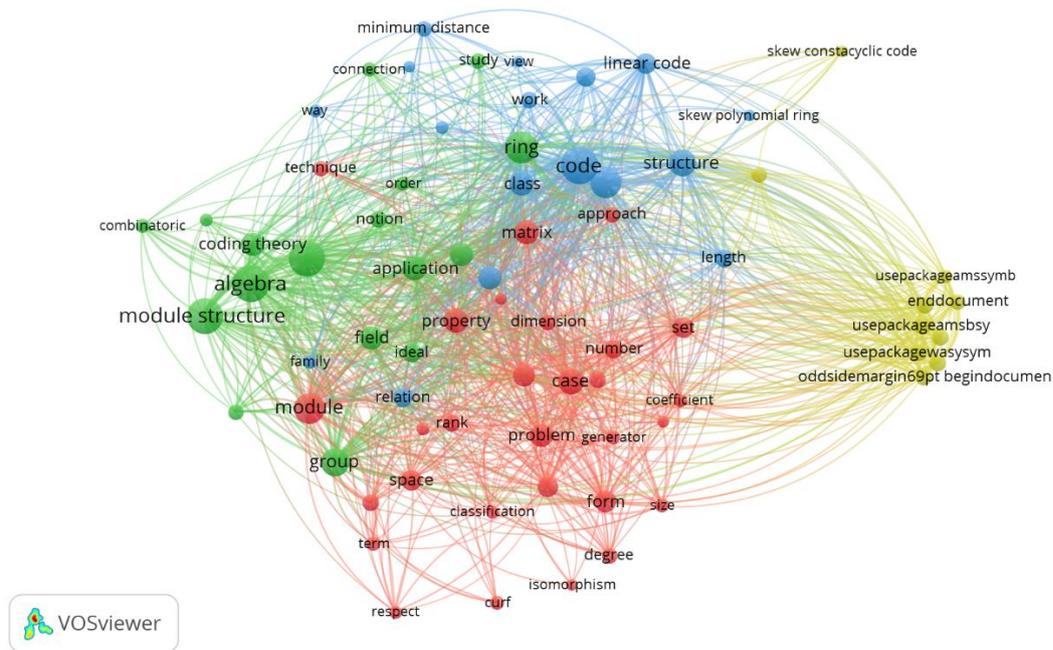

Figure 2. Research topics mapping using VOSviewer

Bibliometrics are the process counting and analyzing many aspects of written communication, as well as the nature and path of development of a field (to the extent that this is reflected through written communication), mathematical and statistical approaches are used to books and other forms of communication [50]. A technique that has been extensively used to evaluate the contributions of research academics in various research disciplines, patterns of publishing, and links between research findings is bibliometric analysis, which was utilized to determine the articles pertinent for the review [51]. The literature review was analyzed using three research indicators: word cocurrence, number of citations, and number of publications [52]. The author using VOSviewer to identify the mapping of the study issue in question, specifically "module structure", "algebra", and "coding theory". Figure 2 shows the bibliometrics of the module structure research topic on coding theory. Figure 2 depicts four hues that rep-resent a cluster of terms that appear often in study, notably red, blue, green, and yellow. The red color indicates keywords that have a strong relationship with the keyword "module". The blue hue represents important terms that have a strong association with the word "code structure" and explain the code's categorization. The green hue indi-cates the link between a term and the keywords sought by the author, which are "module structure", "coding theory", and "algebra". The yellow hue denotes a coding package, which is represented by a latex-shaped file and is interpreted as a "use pack-age" statement. Clusters between colors also suggest a link between the colors red, blue, green, and yellow, but it is weak connection, which means that words in the red cluster seldom appear together with words in the blue, green, and yellow clusters, and vice versa.

The articles obtained from the selection process were explored further to explain about the authors, title, years of research, method used in research, code structures, and the result of the research. Additionally, this analysis of the literature looks at the variations between the fourteen papers evaluated, as well as the benefits and drawbacks of each, and the structure of the code utilized in the form of modules. We learn more about the main results discussed in this article and suggestions for research. Finally, we re-view the research findings from these papers in order to share discoveries in future work, critique the research findings and the application of the code structure, and offer ideas to researchers in order to explain the direction of research development.

**RESULTS AND DISCUSSION**

After going through the article selection procedure stated in the research method, the researcher has a better grasp of the fourteen publications that were chosen. The researcher examined the whole content of the paper, including the author of the article, the year of publication, the title of the study, the number of citations, the method used, the structure of the code mentioned, and the article research results. Further analysis was carried out by the researcher in order to obtain a summary of the data about the article in Table 1.

Table 1. Identification on Articles

| Author, Year of Research | Title | Cited | Method used in research | Code Structures | Result of the research |
|---|---|---|---|---|---|
| Boulagouaz & Leroy, 2013 | $(\sigma, \delta)$-codes | 18 | Pseudo-linear transformations | Cyclic codes, skew cyclic codes, and $(\sigma, \delta)$-codes | The generic matrices and control matrices of $(\sigma, \delta)$-codes, cyclic codes, and skew cyclic codes |
| Gomez-Torrecillas et al., 2017 | Ideal Codes Over Separable Ring Extensions | 14 | Separable extension of rings and ideal codes | Cyclic convolutional codes | A more in-depth grasp of $\sigma$-cyclic convolutional codes including examples and computations in computer programs. |
| Morales, 2016 | On Lee Association Schemes Over $\mathbb{Z}_4$ and Their Terwilliger Algebra | 10 | Terwilliger algebra of Hamming association schemes. | $\mathbb{Z}_4$-codes of length $n$ and Lee association schemes | Proof of the three major theorems of the algebraic structure of $\mathbb{Z}_4$-codes with length $n$. |
| Berger & El Amrani, 2014 | Codes Over Finite Quotients of Polynomial Rings | 7 | Polynomial Euclidean division from the structure of $\mathbb{F}[x]$ | Quasi-cyclic codes | The basis of divisors for quasi-cyclic codes, generalize to the case $\left(\frac{\mathbb{F}[x]}{f(x)}\right)^l$, where $f(x)$ is a monic polynomial, and the canonical generator matrix. |
| Guneri & Ozkaya, 2016 | Multidimensional Quasi-Cyclic and Convolutional Codes | 7 | The Concatenated Structure to show that multidimensional | Multidimensional quasi-cyclic codes | Proof $QnDC$ codes are asymptotically good, conditions of |

| | | | quasi-cyclic codes are asymptotically good. | | noncatastrophicity of encoders, and determining the free distance of convolutional codes. |
|---|---|---|---|---|---|
| Gorla & Ravagnani, 2018 | An Algebraic Framework for End-to-End Physical-Layer Network Coding | 7 | Construction submodule codes and submodule distance | Module codes and submodule codes | Get distribution of information loss and errors in module transmissions and get the minimum-distance decoding in algebraic framework. |
| Xing et al., 2020 | Low-Complexity Chase Decoding of Reed-Solomon Codes Using Module | 7 | Kotter's iterative polynomial construction algorithm, the concept of Grobner basis of a module and Low-Complexity Chase Basis Reduction (LCC-BR) algorithm | Reed-Solomon Codes | Analysis of the low-complexity and low-latency features of LCC-BR algorithm and the numerical results of Reed-Solomon Codes with LCC-BR interpolation technique. |
| Ozbudak & Ozkaya, 2016 | A Minimum Distance Bound for Quasi-$n$D-Cyclic Codes | 6 | The new concatenated structure and the trace representation. | Multidimensional quasi-cyclic (QnDC) codes and nD convolutional codes | A minimum distance bound for multidimensional quasi-cyclic codes and a lower bound on the free distance of certain 2D convolutional codes. |
| Shi et al., 2017 | A Special Class of Quasi-Cyclic Codes | 5 | The construction of module structure of quasi-cyclic codes | The special class of quasi-cyclic codes | Obtain the image of cyclic codes over an extension field with a basis and the theory of special class of quasi-cyclic codes |
| Borello & Willems, 2022 | On The Algebraic Structure of Quasi-Group Codes | 4 | The concept of algebraic description of quasi-cyclic codes and the full group of permutation | Quasi group codes | Get an intrinsic description of quasi group codes of their permutation automorphism groups and some structural properties. |

| Gomez-Torrecillas et al., 2019 | Some Remarks on Non Projective Frobenius Algebras and Linear Codes | 3 | The concept of bilinear forms, finite ring, and MacWilliams identities | Non Projective Frobenius Algebras and Linear Codes | Get a code with a frobenius alphabet and a literature of non-projective Frobenius Algebras. |
|---|---|---|---|---|---|
| Westerback, 2017 | Parity Check Systems of Nonlinear Codes Over Finite Commutative Frobenius Rings | 1 | The concept of parity check systems, Fourier analysis, and MacWilliams identities | Non linear codes over commutative Frobenius rings | The connection between parity check systems and codes and get a minimal distance distribution of codes. |
| Garcia-Rubira & Lopez-Ramos, 2013 | Tensor Products of Ideal Codes Over Hopf Algebras | 0 | Tensor product of simple codes and The Hamming distance | Indecomposable ideal codes in Radford Hopf algebras | Get a behavior of tensor products of indecomposable ideal codes in Radford Hopf algebras. |
| Dyshko & Wood, 2021 | MacWilliams Extension Property for Arbitrary Weights On Linear Codes Over Module Alphabets | 0 | MacWilliams extensions property, The Hamming weight, and the Fourier Transform | Linear codes over module alphabets | Get special matrices of linear codes and the symmetrized weight compositions of linear codes |

The first article [42] was written by Boulagouz and Leroy in 2013, earned 18 citations by other researchers. The method used to answer the aforementioned issues is to apply the concept of pseudo-linear transformation to cyclic codes and more broadly, $(\sigma,\delta)$-codes. The transformation idea is used to achieve research results in the form of the construction of generic matrices and control matrices for cyclic codes, skew cyclic codes, and $(\sigma,\delta)$-codes. This article consists of four sections, which are Introduction and Preeliminaries, Polynomial and pseudo-linear maps, Generic and control matrices of $(\sigma,\delta)$-codes, and $(\sigma,\delta)$-W-codes where W is a polynomial. This article offers various definitions, theorems, and lemmas that are used to help readers grasp new theorems or lemmas. This article also includes examples from each section of the research to help readers comprehend it better because of the practice in the terminology discussed. This article, in general, is readily comprehended by readers, particularly those who are not involved in mathematics, thus it is not unexpected that the contents of this study are cited by many other researchers.

The second article [6] was written by Gomez-Torrecillas et al. in 2017, earned 14 citations by other researchers. The method used to answer the aforementioned issues is to apply the concept of separable extension to rings and ideal codes with one example code that is cyclic convolutional codes. The notion of separable extension will be utilized to address the issues raised in this article, namely a better understanding of cyclic convolutional codes through definition, examples, and computations in computer programming. This article consists of four sections, which are Introduction, Separability and Ore Extensions, Ideal Codes Generated by Idempotents, and Computation of The Idempotent Generator. Section 1 is an introduction of the formation of thoughts regarding this subject. Section 2 goes through the definitions and lemmas of separable extensions, as well as their applications in section 3 and 4. Section 3 outlines the code that will be utilized in this study. Section 4 addresses calculating the code using the principles from Section 2, and the Chinese Remainder Theorem. Overall, this page contains a comprehensive description and discussion of the topic. The goal of this work is also in line with the findings of the research described. However, this research does not include comprehensive illustrations of each theorem or lemma. This study does not include numerical examples, thus readers who are not familiar with mathematics will struggle to grasp the examples provided by the researchers. It is preferable to include several examples in the form of numbers in this research so that the audience may readily understand it.

The third article [47] was written by Morales in 2016, earned 10 citations by other researchers. The method used to answer the aforementioned issues is to apply the concept of Terwilliger algebra of Hamming association schemes. This notion will be extended to the Lee Association Schemes, with research results in the form of demonstrating the three main theorems concerning the algebraic structure of $Z\_4$-codes of with length n. This article consists of five sections, which are Introduction, Preeliminaries, Specht Modules, sl(V)-modules, and main result. Section 1 and 2 outline the roots of the idea for this article as well as the scientific research utilized to resolve the article's primary difficulties. Section 3 provides a quick overview of symmetric group representation theory and recalls several well-known Specht module features. Section 4 review about several key results in representation theory, focusing on the relationship between Specht modules and irreducible sl\_m (C)-modules. In addition, Section 4 characterize irreducible sl\_m (C)-modules from the standpoints of highest weight theory and Weyl modules. Section 5 demonstrates our key findings, namely Theorem of unital algebra homomorphism, Theorem of decomposition into irreducible modules, and Theorem of permutation algebra. Overall, this article contains a variety of definitions, theorems, and lemmas, as well as thorough proofs, making it simple to grasp the sources of these theorems or lemmas. However, there are no examples of theorems or lemmas in this work. It is important to include examples of theorems or lemmas in this study so that readers who are inexperienced with mathematics may grasp it. Another idea is to broaden the code employed in this study, which is like the $Z\_8$-codes of length n.

The fourth article [8] was written by Berger and El Amrani in 2014, earned 7 citations by other researchers. The method used to answer the aforementioned issues is to apply the concept of Polynomial Euclidean division from the structure of F[x]. The polynomial notion is utilized to get research outcomes, namely identifying the generalization of the quasi-cyclic codes and the code's canonical generator matrix. This article consists of five sections, which are Introduction, Codes over A, A canonical generator matrix for A-codes, q-ary image, and conclusion. Section 0 describes the beginnings of the study, the methodologies employed, the research objectives, and the findings from the researcher's analysis. Section 1 provides definitions and basic findings for codes in E as A-modules. Section 2 presents the major result of this study, which is a canonical system of A-code generator. As a result, there is a canonical "generator matrix" for A-codes. This canonical generator matrix is connected to the Hermite form of matrices and may be thought of as an extension of the canonical form for quasi-cyclic codes. Section 3 is devoted to the investigation of the q-ary image of an A-code formed by matching a polynomial with its list of coefficients. One of the most intriguing aspects is that the F-duality of the q-ary image does not correspond with the A-duality even up to permutation, as in the case of cyclic and quasi-cyclic codes. In reality, in general, the F-dual of the q-ary image of an A-code is not the q-ary image of an A-code. Section 4 describes about the conclusion of this findings of the study. This study contains several definitions, theorems, lemmas, and examples for each of the terms introduced. Numerical examples have also been extensively covered here. Algorithm for issue resolution is explained coherently and segregated in distinct tables to help readers grasp the flow of this study. This study is quite good and understandable to the general audience.

The fifth article [46] was written by Guneri and Ozkaya in 2016, earned 7 citations by other researchers. The method used to answer the aforementioned issues is to apply the Concatenated Structure to show that multidimensional quasi-cyclic codes are asymptotically good and use the concept of convolutional codes to find a minimum distance of multidimensional quasi-cyclic codes. The study's findings include demonstrating the asymptotically good of multidimensional quasi-cyclic codes, establishing the non-catastrophicity requirement for the encoder code, and determining the minimal distance from both convolutional codes and the multidimensional quasi-cyclic codes. This article consists of nine sections, which are Introduction, Background on QC (Quasi-Cyclic) and 2D cyclic codes, Quasi 2D cyclic and 3D cyclic codes, Multidimensional QC codes and their Concatenated Structure, QnDC codes are asymptotically good, Connection to Convolutional Codes, Noncatastrophicity of encoders of rank one convolutional codes, A free distance bound for rank one 2D convolutional codes, and Examples. Section 1 describes the introduction of quasi-cyclic codes and multidimensional quasi-cyclic codes in outline and describes the state of the art of research formation in this paper. Section 2 introduces a family of block codes related to QC codes, multidimensional cyclic codes, and multidimensional convolutional codes. Section 3 goes through the quasi-cyclic codes for n=2 and n=3 situations, namely in 2D and 3D form. Section 4 discusses quasi-cyclic codes with n, hereinafter known as multidimensional quasi-cyclic codes. Section 5 demonstrates that employing a concatenated structure, QnDC (multidimensional quasi-cyclic codes) are asymptotically good. Section 6 discusses the link between quasi-cyclic codes and convolutional codes. Section 7 discusses the noncatastrophicity of convolutional encoder codes. Section 8 discusses the convolutional code's shortest minimum distance. Section 9 contains code examples and numerical examples of the explanations in section 1-8. Overall, this article contains several definitions, theorems, and lemmas relating to this topic. Numerical examples are also included to help readers grasp the overall content of the article. This document has 28 pages, which is too much to read in its entirety. Suggestions for this research include reducing the amount of pages and definitions that are unnecessary so that readers comprehend only what is necessary from the research topic.

The sixth article [40] was written by Gorla and Ravagnani in 2018, earned 7 citations by other researchers. The method used to answer the aforementioned issues is to apply the concept of module codes, submodule codes, and submodule distance to get the minimum distance of codes in algebraic network. The approach is used to generate the submodule codes and its minimal distance in order to generate succinct encoding messages and decrease message transmission errors. This article consists of six sections, which are Introduction, Algebraic preliminaries, Submodule codes and submodule distance, Recovering known encoding and decoding schemes, Bounds, and Constructions. Section 0 describes the roots of this paper's topic and includes state of the art research from earlier researchers with the same study emphasis. Section 1 reviews some definitions and findings concerning PIR's (Principal Ideal Ring), modules, length, and row-echelon forms of matrices over PIR's. Section 2 describes submodule codes and distance, as well as how they related to information loss and mistakes in module transmissions. They also illustrate how to compute the submodule distance quickly. Section 3 demonstrates that, in some instances, error-trapping decoding may be understood as a minimum-distance decoding in our paradigm. Section 4 is devoted to submodule code cardinality limits. Section 5 is dedicated to submodule code constructions and codes over ring products. This article contains several definitions, theorems, and lemmas. The example provided each sub-chapter are also adequate for viewers to grasp the overall substance of the article. A few ideas for this article should be able to limit the amount of pages in the paper so that readers don't feel bored from reading too much.

The seventh article [34] was written by Xing et al. in 2020, earned 7 citations by other researchers. The method used to answer the aforementioned issues is to apply the concept of Kotter's iterative polynomial construction algorithm, Grobner basis of a module and Low-Complexity Chase Basis Reduction (LCC-BR) algorithm. With the LCC-BR method, these notions are employed to address the problem of low-complexity and low-latency features. This article consists of eight sections, which are Introduction, Background knowledge, The ACD-MM algorithm, The LCC-BR algorithm, The progressive LCC-BR algorithm, Decoding complexity and latency, Decoding performance, and Conclusion. Section 1 introduces the Reed-Solomon code, the previous researcher's algorithm, and the LCC-BR algorithm. Section 2 explains the theoretical foundation utilized in the development of this article, including definitions of various obscure codes. Section 3 discusses the basic construction, reduction, and transformation of the ACD-MM algorithm (Algebraic Chase Decoding – Module Minimisation). Section 4 discusses the basic construction and working steps of the LCC-BR algorithm. Section 5 discusses the sequence of vector test and the algorithm for the progressive LCC-BR. Section 6 discusses the decoding complexity and latency of the LCC-BR algorithms and the progressive LCC-BR algorithms. Section 7 runs a decoding test simulation on the two algorithms to check if the performance improves and becomes quicker, or if the opposite is true. Section 8 summarizes the findings of the entire research, namely that the LCC-BR algorithm is not overly complicated, making calculation easier, but the progressive LCC-BR method is complex but produces greater message quality than the LCC-BR algorithm. Overall, this article is simple yet covers the entire point of the contents. Articles provide definitions of numerous new concepts introduced, as well as explanations with tables and figures. This article also includes a flowchart to help you understand the direction of the two algorithms covered in this article. A little idea for this research would be to provide examples with each code produced so that readers may better grasp the complete material of this article.

The eighth article [38] was written by Ozbudak and Ozkaya in 2016, earned 6 citations by other researchers. The method used to answer the aforementioned issues is to apply the concept of the new concatenated structure and the trace representation. This concept may be utilized to solve research problems such as calculating the minimum distance bound for the QnDC codes and the lower bound for the 2D convolutional codes. This article consists of six sections, which are Introduction, Background, A new concatenated structure and trace representation of QnDC codes, The minimum distance bound, Examples, and Applications. Section 1 discusses the state of the art in quasi-cyclic codes research as well as the origins of the link between quasi-cyclic codes and convolutional codes. Section 2 defines quasi-cyclic codes and cyclic codes of length 1 and length n, as well as theorems that may be utilized to solve research issues. Section 3 describes the use of the new concatenated structures and trace representations in multidimensional quasi-cyclic codes. Section 4 computes the minimal distance bounds of 2-dimensional and n-dimensional quasi-cyclic code representations. Section 5 gives an example of a quasi-cyclic code and the minimal spacing that may be achieved by plugging in numbers. Section 6 applies the preceding section's computations to convolutional code applications. This article contains a variety of definitions, theorems, visualizations, examples, and their applications to a code structure. A little suggestion for this article is to limit the number of pages and delete stuff that is not required for research purposes, so that readers may be more succinct and comprehend the main essence of the article's contents.

The nineth article [39] was written by Shi et al. in 2017, earned 5 citations by other researchers. The method used to answer the aforementioned issues is to apply the construction of module structure and the concept of basis theory. We gain a specific class of quasi-cyclic codes, as well as their structures and numerical examples, from these construct and notions. This article consists of four sections, which are Introduction, Module structure of quasi-cyclic codes, Numerics, and Conclusion. Section 1 discusses quasi-cyclic codes and the state of the art in earlier research. Section 2 defines the module structure of quasi-cyclic codes and provides evidence for theorems. Section 3 offers a numerical example of a special class quasi-cyclic codes employing the MDS Reed-Solomon codes. Section 4 presents the study's conclusion, which is that a full class of quasi-cyclic codes has been developed. This article contains definitions and theorems in general, although the definitions covered in this research are too brief and lack solid theoretical studies. Furthermore, the numerical examples are self-explanatory and simple to grasp. Suggestions for this research include adding to the bibliography to reinforce the theoretical works covered and adding definitions and theorems to strengthen the researchers' opinions.

The tenth article [12] was written by Borello and Willems in 2022, earned 4 citations by other researchers. The method used to answer the aforementioned issues is to apply the concept of algebraic description of quasi-cyclic codes and the full group of permutation. With these two concepts, this study obtained research results, namely obtaining an intrinsic description of quasi group codes and some of the structures of the code. This article consists of five sections, which are Introduction, Background, Codes with a free acting group of symmetries, The concatenated structure of quasi group codes, and Self-duality of quasi group codes. Section 1 discusses the genesis of this study, earlier research on other codes, and the progression of this research. Section 2 defines the terms used in this study to answer research questions. Section 3 covers the definitions, theorems, and remarks relating to the codes for symmetric groups. Section 4 discusses the concatenated structure of quasi group codes creation. Section 5 discusses the duality of quasi group codes. Overall, this article contains definitions, theorems, and remarks that support the theoretical foundation of this research. Furthermore, numerical examples that are simply comprehended by the audience have been provided, and the reference is sufficient to ensure the credibility of this research.

The eleventh article [44] was written by Gomez-Torrecillas et al. in 2019, earned 3 citations by other researchers. The method used to answer the aforementioned issues is to apply the concept of bilinear forms, finite ring, and MacWilliams identities. This concept will be utilized subsequently to solve research problems, such as building codes using the Frobenius alphabet and examining the literature. This article consists of six sections, which are Introduction, Preliminaries on bilinear forms and annihilators, Non projective Frobenius algebras, Annihilators in non projective Frobenius extensions, Non projective Frobenius algebras and Frobenius rings, and Codes with a Frobenius alphabet. Section 1 provides the research hypothesis, the flow of the research, and an explanation of the sections that evolved from the investigation. Section 2 contains claims and proofs of preliminary conclusions on balanced (also known as associative) bilinear forms over bimodules; we only present proofs that we thought instructive. Section 3 contains our proposed definition of a non projective Frobenius algebra over a commutative ring, as well as numerous similar characterizations; as previously stated, projectivity of the algebra over the base ring is not essential. One of these analogous criteria is the existence of a Frobenius functional, which will be found to generalize and perform a similar role to the generating character. Section 4 extends results from a finite Frobenius ring to a non projective Frobenius algebra on annihilators associated with a non degenerate bilinear form. Section 5 contains our finding that if an algebra R is finitely produced as a K-module over a Frobenius commutative ring K, then R is a non projective Frobenius algebra over K if and only if R is a Frobenius ring. This is especially true for finite rings when considered as algebras over their characteristic subrings. We also present a method for creating new Frobenius algebras from existing ones that is based on skew polynomials. Section 6 describes how to build new finite Frobenius rings from old ones, as well as how the general findings on bilinear forms defined on modules over non projective Frobenius algebras discovered in the preceding sections may be used to obtain the major conditions. Overall, this article contains definitions, lemmas, corollaries, and remarks on the research theory. Furthermore, additional instances are provided toward the end of the section. Suggestions for this research include increasing the quantity of numerical examples so that viewers may better grasp the complete contents of the article and include references so that the researcher's ideas have credibility.

The twelfth article [48] was written by Westerback in 2017, earned 1 citation by other researchers. The method used to answer the aforementioned issues is to apply the concept of parity check systems, Fourier analysis, and MacWilliams identities. These three concepts are employed to generate research findings, namely the relationship between the parity check system and the code, followed by the minimal distance distribution obtained from the code. This article consists of six sections, which are Introduction, Preliminaries, Parity check systems and Codes over R, Fourier analysis on codes over finite commutative Frobenius rings, and Parity check systems, distance, and Fourier coefficients. Section 1 summarizes past study of this topic, presents the direction of this research, and provides a brief summary of each section's discussion in this article. Section 2 provides some fundamental facts and nomenclature for finite commutative Frobenius rings R, and parity check systems over R. Section 3 discusses the underlying relationship between parity check systems and codes. Section 4's first section introduces some fundamental notions and facts regarding Fourier analysis on finite Abelian groups. In the second part, some Fourier analysis results on codes over R are presented. Section 5 largely deals with how to achieve the minimal distance and distance distribution of codes via the use of parity check systems, and how parity check systems are related to characters. This article contains definitions, theorems, propositions, and examples. Many numerical examples are included in the examples, which can help viewers comprehend. Suggestions for this research include reducing the amount of pages in the paper so that readers do not encounter too many foreign phrases and can grasp the research's main aim.

The thirteenth articles [49] were written by Garcia-Rubira and Lopez-Ramos in 2013, not getting a citation from other researchers. The method used to answer the aforementioned issues is to apply the concept of tensor product of simple codes and the Hamming distance. The idea is used to derive the behavior of ideal codes that are not decomposed in Radford Hopf algebras. This article consists of four sections, which are Introduction, Two families of Hopf algebras, Tensor product of indecomposables, and Some practical considerations. Section 1 explores the roots of this research topic as well as the state of the art of prior researchers. Section 2 reviews the Taft and Radford Hopf algebra definitions. In order to comprehend the nature of the indecomposables in the Radford Hopf algebra family, we outline the proof of the theorem found in Theorem 2.2 that links both families. Section 3 investigates the tensor products of indecomposable codes. In the Taft Hopf algebra, the tensor product of two any indecomposables may be identified with a direct sum of indecomposable ideal codes. In the instances of simple code tensor products, we obtained a semisimple ideal code. When analyzing the family of Radford Hopf algebras, we now separate the fact that semisimplicity of tensor product of simple ideal codes is lost. Section 4 considers some practices of applying indecomposable ideal codes to numerical examples.

The last articles [45] were written by Dyshko and Wood in 2021, not getting a citation from other researchers. The method used to answer the aforementioned issues is to apply the concept of MacWilliams extensions property, the Hamming weight, and the Fourier Transform. These three concepts, namely a particular matrix of linear codes and their symmetrical weight composition, are applied to generate research findings. This article consists of seven sections, which are Introduction, Background on linear codes, modules, and characters, Fourier transform calculations, Defining special matrices, Main theorem, Examples, and Vera Pless: a reflection. Section 1 describes the state of the art in this research and briefly explains the paper's format. Section 2 introduces linear codes, pseudo-injective modules, characters, and the Fourier transform. When A is pseudo-injective, two functionals from M to A have the same kernel if and only if they share the same $GL_R$ (A)-orbit. Section 3 performs Fourier transform computations and discusses its interpretation in terms of transforms ω along pictures of functionals. Section 4 establishes the matrices $Q^{\wedge}S$, whose nondegeneracy is required by the main theorem. Section 5 discusses the main theorem and its proof using a recursive argument. Section 6 gives several instances of how the main theorem might be used in the form of algebraic expressions. Section 7 is a small homage to the previous author who may categorize a code that is extensively utilized in this author's research. This article contains several definitions, conventions, theorems, propositions, lemmas, and corollaries. Each theorem is supported by a proof in this article. The examples provided are likewise numerous and easily comprehended by the audience. A few ideas for this research, ideally minimizing the number of pages that are not required in this research so that readers may comprehend the substance of this article's contents.

## CONCLUSIONS

This systematic literature review was conducted by examining three databases for articles: Google scholar, Dimensions, and Science Direct. The search was conducted by restricting the research year from 2012 to 2022. There were 470 articles returned from the article search. Following the completion of the literature selection process, 14 articles were selected for further investigation of the huge understanding of the article. The terms module structure, coding theory, and algebra were used to search for articles. The majority of the articles received are good since they include definitions, theorems and their proofs, lemmas, and numerical examples. The articles received are relevant to the topics sought, thus comments from researchers are limited to the structure of the article's contents. The consideration for further research is to further analyze the suggested code structure which is the optimal code structure to address the research problem.

## ACKNOWLEDGMENTS

This work was supported by Hibah Riset Unpad Riset Percepatan Lektor Kepala Contract Number 1549/UN6.3.1/PT.00/2023.

## REFERENCES

[1]   P. L. Clark, "COMMUTATIVE ALGEBRA."

[2]   F. Henglein, R. Kaarsgaard, and M. K. Mathiesen, "The Programming of Algebra," in *Electronic Proceedings in Theoretical Computer Science, EPTCS*, Jun. 2022, vol. 360, pp. 71–92. doi: 10.4204/EPTCS.360.4.

[3]   T. Westerbäck, "Parity check systems of nonlinear codes over finite commutative Frobenius rings," Apr. 2016, [Online]. Available: http://arxiv.org/abs/1604.06996

[4]   A. K. Sinha, "Application of Module Structure of Algebra in Coding Theory in Different Branches of Engineering," 2013.

[5]   E. Gorla and A. Ravagnani, "An algebraic framework for end-to-end physical-layer network coding," Nov. 2016, [Online]. Available: http://arxiv.org/abs/1611.04226

[6]   J. Gómez-Torrecillas, F. J. Lobillo, and G. Navarro, "Ideal codes over separable ring extensions," Aug. 2014, [Online]. Available: http://arxiv.org/abs/1408.1546

[7]   M. Sala, T. Mora, L. Perret, S. Sakata, and C. Traverso, *Gröbner bases, coding, and cryptography*. Springer Berlin Heidelberg, 2009. doi: 10.1007/978-3-540-93806-4.

[8]   T. P. Berger and N. el Amrani, "Codes over finite quotients of polynomial rings," *Finite Fields and their Applications*, vol. 25, pp. 165–181, 2014, doi: 10.1016/j.ffa.2013.09.004.

[9]   J. A. Wood, "CODE EQUIVALENCE CHARACTERIZES FINITE FROBENIUS RINGS," 2008. [Online]. Available: https://www.ams.org/journal-terms-of-use

[10]  S. Mahmoudi, F. Mirmohammadirad, and K. Samei, "Fq-linear codes over Fq-algebras," *Finite Fields and their Applications*, vol. 64, Jun. 2020, doi: 10.1016/j.ffa.2020.101665.

[11]  S. Ling and P. Solé, "On the Algebraic Structure of Quasi-Cyclic Codes I: Finite Fields," 2001.

[12]  M. Borello and W. Willems, "On the algebraic structure of quasi group codes," Dec. 2019, [Online]. Available: http://arxiv.org/abs/1912.09167

[13]  S. Ling and P. Solé, "On the algebraic structure of quasi-cyclic codes III: Generator theory," *IEEE Trans Inf Theory*, vol. 51, no. 7, pp. 2692–2700, Jul. 2005, doi: 10.1109/TIT.2005.850142.

[14]  D. Boucher, W. Geiselmann, and F. Ulmer, "Skew-cyclic codes," *Applicable Algebra in Engineering, Communications and Computing*, vol. 18, no. 4, pp. 379–389, 2007, doi: 10.1007/s00200-007-0043-z.

[15]  D. Boucher, P. Sole, F. Ulmer, D. Boucher, P. Solé, and F. Ulmer, "Skew Constacyclic Codes over Galois Rings," 2008. [Online]. Available: https://hal.archives-ouvertes.fr/hal-00359833

[16]  J. Cuadra, J. M. García-Rubira, and J. A. López-Ramos, "Determining all indecomposable codes over some Hopf algebras," *J Comput Appl Math*, vol. 235, no. 7, pp. 1833–1839, Feb. 2011, doi: 10.1016/j.cam.2010.07.014.

[17]  A. Edelmayer and International Symposium on the Mathematical Theory of Networks and Systems., *MTNS 2010 : 19th international symposium on mathematical theory of networks and systems, MTNS 2010, Budapest, Hungary, 05-09.07.2010*. 2010.

[18]  D. Manchon, "Hopf Algebras in Renormalisation," *Handbook of Algebra*, vol. 5. pp. 365–427, 2008. doi: 10.1016/S1570-7954(07)05007-3.

[19]  J. J. Climent, D. Napp, C. Perea, and R. Pinto, "A construction of MDS 2D convolutional codes of rate 1 / n based on superregular matrices," *Linear Algebra Appl*, vol. 437, no. 3, pp. 766–780, Aug. 2012, doi: 10.1016/j.laa.2012.02.032.

[20]  H. Gluesing-Luerssen, J. Rosenthal, and P. A. Weiner, "Duality between Multidimensional Convolutional Codes and Systems *," 1999.

[21]  K. Lally, "Algebraic lower bounds on the free distance of convolutional codes," *IEEE Trans Inf Theory*, vol. 52, no. 5, pp. 2101–2110, May 2006, doi: 10.1109/TIT.2006.872980.

[22]  Q. T. Sun, J. Yuan, T. Huang, and K. W. Shum, "Lattice network codes based on eisenstein integers," *IEEE Transactions on Communications*, vol. 61, no. 7, pp. 2713–2725, 2013, doi: 10.1109/TCOMM.2013.050813.120759.

[23]  T. Y. Lam and A. Leroy, "Wedderburn polynomials over division rings, I," *J Pure Appl Algebra*, vol. 186, no. 1, pp. 43–76, Jan. 2004, doi: 10.1016/S0022-4049(03)00125-7.

[24]    T. Y. Lam, A. Leroy, and A. Ozturk, "Wedderburn polynomials over division rings, II," Jun. 2007, [Online]. Available: http://arxiv.org/abs/0706.3515

[25]    T. Y. Lam and A. Leroy, "Wedderburn Polynomials over Division Rings," 1999.

[26]    V. Guruswami and M. Sudan, "Improved Decoding of Reed-Solomon and Algebraic-Geometry Codes," 1999.

[27]    R. Koetter and A. Vardy, "Algebraic Soft-Decision Decoding of Reed-Solomon Codes," *IEEE Trans Inf Theory*, vol. 49, no. 11, pp. 2809–2825, Nov. 2003, doi: 10.1109/TIT.2003.819332.

[28]    R. Koetter and A. Vardy, "A Complexity Reducing Transformation in Algebraic List Decoding of Reed-Solomon Codes."

[29]    R. Koetter, J. Ma, and A. Vardy, "The Re-Encoding Transformation in Algebraic List-Decoding of Reed-Solomon Codes," May 2010, [Online]. Available: http://arxiv.org/abs/1005.5734

[30]    K. Lee and M. E. O'Sullivan, "List decoding of Reed-Solomon codes from a Gröbner basis perspective," *J Symb Comput*, vol. 43, no. 9, pp. 645–658, Sep. 2008, doi: 10.1016/j.jsc.2008.01.002.

[31]    K. Lee and M. E. O'sullivan, "An Interpolation Algorithm using Gröbner Bases for Soft-Decision Decoding of Reed-Solomon Codes."

[32]    M. Alekhnovich, "Linear Diophantine equations over polynomials and soft decoding of Reed-Solomon codes," 2002.

[33]    A. Zeh, C. Gentner, and D. Augot, "An Interpolation Procedure for List Decoding Reed--Solomon codes Based on Generalized Key Equations," Oct. 2011, doi: 10.1109/TIT.2011.2162160.

[34]    J. Xing, L. Chen, and M. Bossert, "Low-Complexity Chase Decoding of Reed-Solomon Codes Using Module," *IEEE Transactions on Communications*, vol. 68, no. 10, pp. 6012–6022, Oct. 2020, doi: 10.1109/TCOMM.2020.3011991.

[35]    J. Xing, L. Chen, and M. Bossert, "Progressive Algebraic Soft-Decision Decoding of Reed-Solomon Codes Using Module Minimization," in *IEEE Transactions on Communications*, Nov. 2019, vol. 67, no. 11, pp. 7379–7391. doi: 10.1109/TCOMM.2019.2927207.

[36]    L. Chen, S. Tang, and X. Ma, "Progressive algebraic soft-decision decoding of reed-solomon codes," *IEEE Transactions on Communications*, vol. 61, no. 2, pp. 433–442, 2013, doi: 10.1109/TCOMM.2012.100912.110752.

[37]    L. Chen and M. Bossert, "Algebraic Chase Decoding of Reed-Solomon Codes Using Module Minimisation," 2016.

[38]    F. Özbudak and B. Özkaya, "A minimum distance bound for quasi-nD-cyclic codes," *Finite Fields and their Applications*, vol. 41, pp. 193–222, Sep. 2016, doi: 10.1016/j.ffa.2016.06.004.

[39]    M. Shi, J. Tang, M. Ge, L. Sok, and P. Solé, "A special class of quasi-cyclic codes," *Bull Aust Math Soc*, vol. 96, no. 3, pp. 513–518, Dec. 2017, doi: 10.1017/S0004972717000636.

[40]    E. Gorla and A. Ravagnani, "An algebraic framework for end-to-end physical-layer network coding," Nov. 2016, [Online]. Available: http://arxiv.org/abs/1611.04226

[41]    T. P. Berger and N. el Amrani, "Codes over finite quotients of polynomial rings," *Finite Fields and their Applications*, vol. 25, pp. 165–181, 2014, doi: 10.1016/j.ffa.2013.09.004.

[42]    M. Boulagouaz and A. Leroy, "(Sigma-Delta) Codes," Apr. 2013, [Online]. Available: http://arxiv.org/abs/1304.6518

[43]    J. Gómez-Torrecillas, F. J. Lobillo, and G. Navarro, "Ideal codes over separable ring extensions," Aug. 2014, [Online]. Available: http://arxiv.org/abs/1408.1546

[44]    J. Gómez-Torrecillas, E. Hieta-Aho, F. J. Lobillo, S. López-Permouth, and G. Navarro, "Some remarks on non projective Frobenius algebras and linear codes," *Des Codes Cryptogr*, vol. 88, no. 1, pp. 1–15, Jan. 2020, doi: 10.1007/s10623-019-00666-1.

[45]    S. Dyshko and J. A. Wood, "MacWilliams extension property for arbitrary weights on linear codes over module alphabets," *Des Codes Cryptogr*, vol. 90, no. 11, pp. 2683–2701, Nov. 2022, doi: 10.1007/s10623-021-00945-w.

[46]    C. Guneri and B. Ozkaya, "Multidimensional quasi-cyclic and convolutional codes," *IEEE Trans Inf Theory*, vol. 62, no. 12, pp. 6772–6785, Dec. 2016, doi: 10.1109/TIT.2016.2616467.

[47]    J. V. S. Morales, "On Lee association schemes over Z4 and their Terwilliger algebra," *Linear Algebra Appl*, vol. 510, pp. 311–328, Dec. 2016, doi: 10.1016/j.laa.2016.08.033.

[48]    T. Westerbäck, "Parity check systems of nonlinear codes over finite commutative Frobenius rings," Apr. 2016, [Online]. Available: http://arxiv.org/abs/1604.06996

[49]    J. M. García-Rubira and J. A. López-Ramos, "Tensor products of ideal codes over Hopf algebras," *Sci China Math*, vol. 56, no. 4, pp. 737–744, Apr. 2013, doi: 10.1007/s11425-013-4579-z.

[50]    T. Shanmugam and S. Thanuskodi, "Bibliometric Analysis of the Indian Journal of Chemistry, S. Thanuskodi," *Bibliometric Analysis of the Indian Journal of Chemistry*, [Online]. Available: http://unllib.unl.edu/LPP/

[51]    K. C. Garg and C. Sharma, "Bibliometrics of library and information science research in India during 2004-2015," *DESIDOC Journal of Library and Information Technology*, vol. 37, no. 3, pp. 221–227, 2017, doi: 10.14429/djlit.37.3.11188.

[52]    H. Mallawaarachchi, Y. Sandanayake, G. Karunasena, and C. Liu, "Unveiling the conceptual development of industrial symbiosis: Bibliometric analysis," *Journal of Cleaner Production*, vol. 258. Elsevier Ltd, Jun. 10, 2020. doi: 10.1016/j.jclepro.2020.120618.

      1.